\documentclass[twocolumn,aps,pra,nofootinbib,groupedaddress,amsmath,amssymb]{revtex4-2}
\usepackage{graphicx}
\usepackage[dvipsnames]{xcolor}
\usepackage{bm}
\usepackage[colorlinks=true,linkcolor=blue,urlcolor=blue,anchorcolor=blue,citecolor=blue,bookmarksnumbered]{hyperref}
\usepackage{float}
\usepackage{times}

\begin{document}
\title{Multipolar Fermi-surface deformation in a Rydberg-dressed Fermi gas with long-range anisotropic interactions}
\author{Yijia Zhou$^{1,2}$, Rejish Nath$^3$, Haibin Wu$^{4,5}$, Igor Lesanovsky$^{1,6}$, Weibin Li$^1$}
\affiliation{$^1$School of Physics and Astronomy and Centre for the Mathematics and Theoretical Physics of Quantum Non-equilibrium Systems, University of Nottingham, Nottingham NG7 2RD, United Kingdom\\
$^2$Graduate School of China Academy of Engineering Physics, Beijing 100193, China \\ 
$^3$Department of Physics, Indian Institute of Science Education and Research, Dr. Homi Bhabha Road, Pune 411008, Maharashtra, India\\
$^4$State Key Laboratory of Precision Spectroscopy, East China Normal University, Shanghai 200062, China \\
$^5$Collaborative Innovation Center of Extreme Optics, Shanxi University, Taiyuan 030006, China\\
$^6$Institut f\"ur Theoretische Physik, University of T\"ubingen, 72076 T\"ubingen, Germany}
\begin{abstract}
We study theoretically the deformation of the Fermi surface (FS) of a three-dimensional gas of Rydberg-dressed $^6$Li atoms. The laser dressing to high-lying Rydberg $D$ states results in angle-dependent soft-core-shaped interactions whose anisotropy is described by multiple spherical harmonics. We show that this can drastically modify the shape of the FS and that its deformation depends on the interplay between the Fermi momentum $k_F$ and the reciprocal momentum $\bar{k}$ corresponding to the characteristic soft-core radius of the dressing-induced potential. When $k_F< \bar{k}$, the dressed interaction stretches a spherical FS into an ellipsoid. When $k_F\gtrsim \bar{k}$, complex deformations are encountered which exhibit multipolar characteristics. We analyze the formation of Cooper pairs around the deformed FS and show that they occupy large orbital angular momentum states ($p$, $f$, and $h$ wave) coherently. Our study demonstrates that Rydberg dressing to high angular momentum states may pave a route toward the investigation of unconventional Fermi gases and multiwave superconductivity.
\end{abstract}
\date{\today}
\keywords{}
\maketitle

\textit{Introduction.}
Fermi surfaces (FSs), describing the occupation of momentum space by fermions, lie at the heart of Fermi liquid theory~\cite{baym_landau_2008}. Though typically being a sphere in free space, FSs can be deformed by anisotropic two-body interactions, resulting in novel physics, manifesting in the nematic phase~\cite{kivelson_electronic_1998} and the Pomeranchuk instability~\cite{Pomeranchuk}. Quantum simulators based on ultracold atoms provide a flexible platform for realizing Fermi gases with controllable FS~\cite{giorgini_theory_2008}. Spherical FSs have been observed in free~\cite{drake_direct_2012} and weakly interacting gases~\cite{mukherjee_homogeneous_2017-1}. The spherical symmetry can be broken by anisotropic dipole-dipole interactions~\cite{marinescu_controlling_1998,ni_high_2008,lu_quantum_2012-1,baranov_condensed_2012,shi_ultracold_2013-1,baier_realization_2018-1,trautmann_dipolar_2018}, and the deformation of the FS from a sphere to ellipsoid~\cite{miyakawa_phase-space_2008,bruun_quantum_2008,fregoso_ferronematic_2009,zhang_fermi_2009,sogo_dynamical_2009,yamaguchi_density-wave_2010,baillie_thermodynamics_2010,zhao_hartree-fock-bogoliubov_2010-1, wachtler_low-lying_2017} has been observed in polarized dipolar gases~\cite{aikawa_observation_2014}. 

A newly emerged approach to creating long-range interactions is so-called Rydberg dressing~\cite{PhysRevA.65.041803, henkel_three-dimensional_2010, pupillo_strongly_2010, PhysRevLett.105.160404, PhysRevLett.105.135301, PhysRevA.85.053615}: Using off-resonant lasers, the strong van der Waals interaction between electronically high-lying Rydberg atoms~\cite{saffman_quantum_2010,PhysRevX.8.021069, leseleuc_observation_2019} is mapped to the ground state, yielding an effective interaction between dressed atoms that possesses a characteristic ``soft-core'' shape. The radius $\bar{R}$ at which the soft core occurs~\cite{henkel_three-dimensional_2010} and the potential shape can be tuned by dressing to different Rydberg states~\cite{PhysRevLett.114.173002, PhysRevLett.114.243002, cinti_defect-induced_2014, PhysRevLett.115.093002, PhysRevA.102.063307}. Generically, the collective behavior of the atomic gas is strongly impacted by the soft-core interaction~\cite{henkel_three-dimensional_2010, PhysRevLett.105.160404, cinti_defect-induced_2014,zhou_quench_2020}.  In particular, the stability of elementary excitation and the emergence of supersolidity is connected to $\bar{R}$. So far experimental and theoretical studies have mostly focused on spin~\cite{zeiher_many-body_2016, PhysRevX.7.041063, PhysRevLett.124.063601} and bosonic systems~\cite{henkel_three-dimensional_2010, pupillo_strongly_2010, PhysRevLett.105.160404, PhysRevLett.105.135301, PhysRevA.97.023619, PhysRevA.85.053615, mukherjee_phase-imprinting_2015, li_many-body_2020}. In a recent experiment~\cite{guardado-sanchez_quench_2020}, dressing of the fermionic $^6$Li atoms to Rydberg $P$ states was demonstrated, which indeed highlighted new opportunities for the exploration of correlated many-body phases with Rydberg-dressed fermions~\cite{xiong_topological_2014-1,li_exotic_2015,Li2016a,PhysRevA.96.053611,keles_f-wave_2020}.

\begin{figure}
	\centering
	\includegraphics[width=0.99\linewidth]{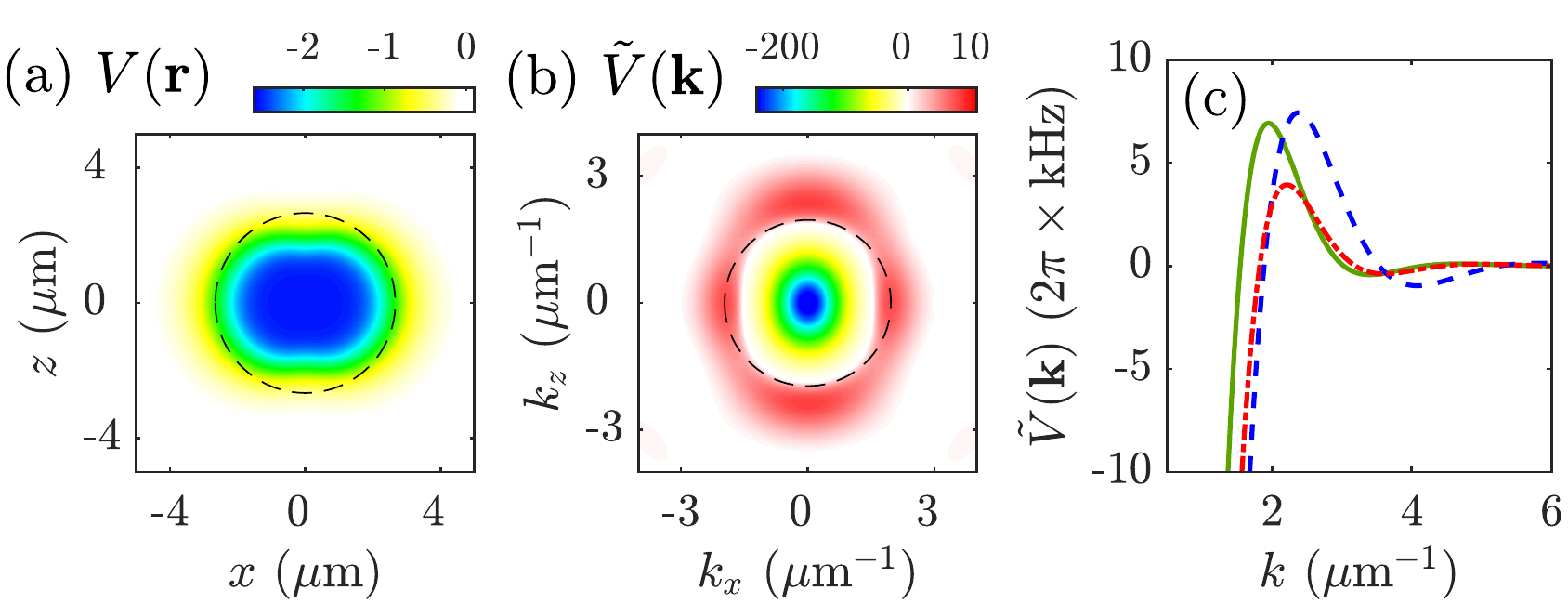}
	\caption{ Anisotropic soft-core interaction through Rydberg-dressing. (a) Angular dependence of the attractive, soft-core interaction $V(d,\theta)$ in the $xz$ plane. The interaction is stronger (weaker) along the $x(z)$ axis. The dashed line shows the circle whose radius is $\bar{R}$. (b) The Fourier transformed potential $\tilde{V}(\mathbf{k})$ is attractive at small momentum and becomes repulsive around $\bar{k}\approx 5\pi/(3\bar{R})$ (dashed line).  (c) Profile of $\tilde{V}(\mathbf{k})$ cuts along the polar angle  $\theta_k = 0$ (blue dashed),  $\pi/2$ (green solid), and $\pi$ (red dash-dot). We consider $n=40$, $V_0\approx - 2\pi\times2.53$~kHz, and $\delta=-2\pi\times40$~MHz.   The energy unit in panels (a) and (b) is $2\pi\times{\rm kHz}$.  See text for details.}
	\label{fig:introduction}
\end{figure}

In this work, we study the FS deformation of a spin-polarized,
zero-temperature gas of $^6$Li atoms through Rydberg dressing. We find that the emerging long-range attractive two-body interaction $V(\mathbf{r})$ combines monopole, dipole, and quadrupole components [Fig.~\ref{fig:introduction}(a)], when dressing the ground-state atoms to a high-lying Rydberg $D$ state. In momentum space, the interaction $\tilde{V}(\mathbf{k})$ is largest at around ${k}(\theta)=5\pi/3R(\theta)$ with $R(\theta)$ being the angle-dependent soft-core radius [Fig.~\ref{fig:introduction}(b)]. When the Fermi momentum becomes larger than the minimum ${k}(\theta)$, the FS is strongly deformed and gains multipolar symmetries. We show that this deformation is accompanied by the formation of Cooper pairs, where $p$-, $f$-, and $h$-wave pairing coexist coherently when multipolar FS deformation occurs. The $p$- and $f$-wave pairing are key to the superfluidity of $^3$He~\cite{israelsson_f-wave_1986, davis_discovery_2008} and unconventional superconductors as shown in Ref.~\cite{stewart_unconventional_2017}, while $h$-wave pairing has, to the best of our knowledge, not been identified in any other system. Our work shows that Rydberg-dressed fermions provide a tunable quantum simulator for investigating novel FS deformation and nontrivial pairing  states.

\begin{figure} [b]
	\centering
	\includegraphics[width=\linewidth]{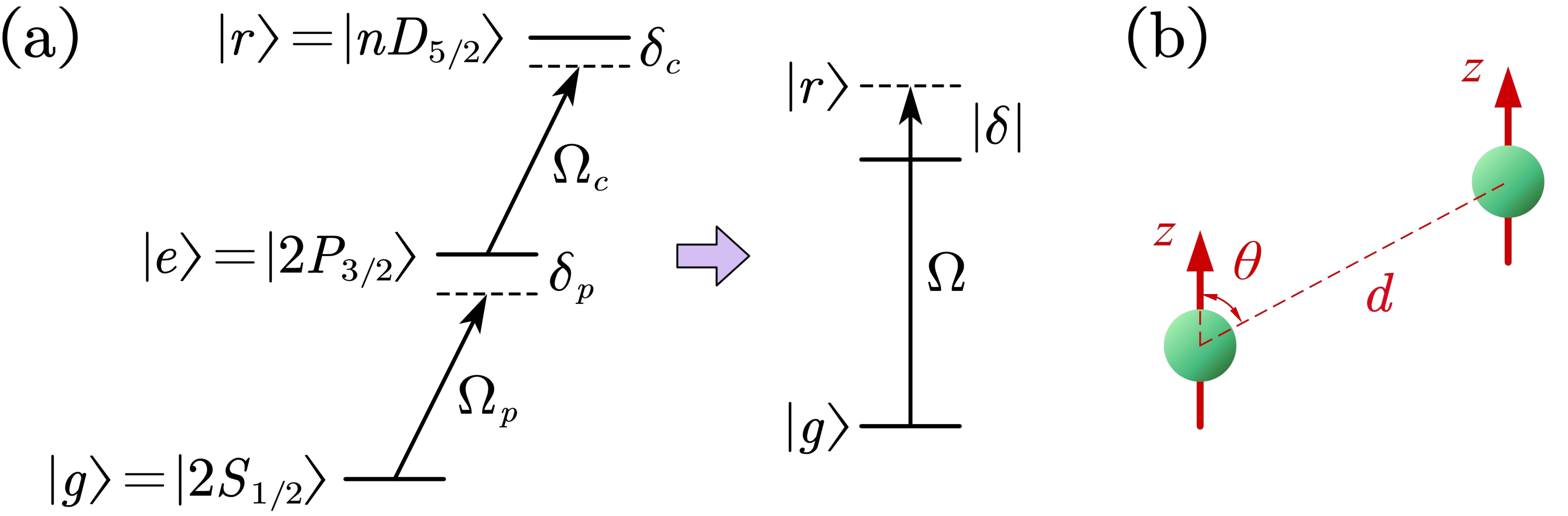}
	\caption{ Rydberg dressing. (a) Two-photon dressing scheme. The probe laser is far and blue detuned while the control laser is on resonance. The three-level system can be simplified to a two-level model by adiabatic elimination. The effective Rabi frequency is $\Omega = \Omega_p\Omega_c / (2\delta_p)$, and the effective detuning is $\delta = \delta_p + \delta_c - (\Omega_c^2-\Omega_p^2)/(4\delta_p)$. (b) Displacement between two atoms, and $\theta$ is the angle between the displacement and quantization axis. }
	\label{fig:diagram}
\end{figure}

\textit{Anisotropic Rydberg-dressed interaction.}
In our setting, each Li$^6$ atom consists of the electronic ground state $|g\rangle = |2S_{1/2}\rangle$, intermediate state $|e\rangle= |2P_{3/2}\rangle$, and Rydberg state $|r\rangle=|nD_{5/2}\rangle$ ($n$ denotes the principal quantum number), as shown in Fig.~\ref{fig:diagram}(a). A probe and a control laser couple the  $|g\rangle \leftrightarrow |e\rangle$ and $|e\rangle \leftrightarrow |r\rangle$ transitions. This coupling is described by the single-atom Hamiltonian $\hat{H}^{(1)} = \delta_p |e\rangle\langle e| + (\delta_p+\delta_c) |r\rangle\langle r| + (\Omega_p/2) ( |g\rangle\langle e| + |e\rangle\langle g| ) + (\Omega_c/2) ( |e\rangle\langle r| + |r\rangle\langle e| )$, where $\Omega_p$ ($\Omega_c$) and $\delta_p$ ($\delta_c$) are Rabi frequency and detuning of the probe (control) laser, respectively.  The state $|e\rangle$ can be adiabatically eliminated, provided that $|\delta_p|\gg |\Omega_p|,\, |\Omega_c|$, and $|\delta_c|$. This leads to an effective Hamiltonian,  $\hat{H}^{(1)}_{\rm e} \approx   \delta |r\rangle\langle r| - \Omega/2 ( |g\rangle\langle r| + |r\rangle\langle g| )$ with the effective detuning $\delta \approx \delta_p + \delta_c - (\Omega_c^2-\Omega_p^2)/(4\delta_p)$, and Rabi frequency $\Omega = \Omega_p\Omega_c/(2\delta_p)$.

When excited to the Rydberg state, two atoms interact through the angular dependent van der Waals (vdW) interaction $V_{\rm vdW}(d,\theta)=C_6(\theta)/d^6$, where $C_6(\theta)$ is the dispersion coefficient,  $d=|\mathbf{r}-\mathbf{r}'|$ is the distance between the atoms (locating at $\mathbf{r}$ and $\mathbf{r}'$), and $\theta$ is the angle between the atoms and quantization $z$ axis [see Fig.~\ref{fig:diagram}(b)]. We obtain the two-atom Hamiltonian, $\hat{H}^{(2)} = \hat{H}_{\rm e}^{(1)}(\mathbf{r}) + \hat{H}_{\rm e}^{(1)}(\mathbf{r}') + V_{\rm vdW}(d,\theta) |rr\rangle\langle rr|$.  In $|nD_{5/2},5/2\rangle$ states of $^6$Li atoms, the dispersion coefficient can be expressed in terms of spherical harmonics $Y_{lm}(\theta)$ ($l$ is an even number)
\begin{equation}
	\label{eq:C6}
	C_6(\theta)=\mathcal{C}_0Y_{00}(\theta) + \mathcal{C}_2 Y_{20}(\theta) + \mathcal{C}_4Y_{40}(\theta),
\end{equation}
which consists of monopole, dipole and quadrupole components~\cite{kamenski_energy_2017} with strength  $[\mathcal{C}_0,\mathcal{C}_2,\mathcal{C}_4]= [-17.32, 5.98, -0.076]\times 10^{-17} n^{11}{\rm GHz\ \mu m^6}$~\cite{SM}. 

When $|\delta|\gg |\Omega|$, the ground-state atoms are weakly coupled to the Rydberg state, such that they experience a weaker, laser-dressed two-body interaction.  The dressed interaction can be obtained through the conventional fourth-order perturbation calculation~\cite{Johnson2010,henkel_three-dimensional_2010}.  
One can also derive the dressed interaction through the projection operator method~\cite{Cohen-Tannoudji_1998}. We first define the two-atom ground-state subspace given by $\hat{P} = |gg\rangle\langle gg|$. The effective Hamiltonian in this subspace is
\begin{equation} \label{eq:PHP}
	\hat{H}_{\rm eff} = \hat{P}\hat{H}\hat{P} - \hat{P}\hat{H}\hat{Q} \frac{1}{\hat{Q}\hat{H}\hat{Q}} \hat{Q}\hat{H}\hat{P},
\end{equation}
where  $\hat{Q}=\hat{\mathbb{I}}-\hat{P}$ with $\hat{\mathbb{I}}$ being the identity operator.
This results to the dressed interaction potential
\begin{equation}
	V(d,\theta) =  V_0 \frac{R^6(\theta)}{d^6+R^6(\theta)},
\end{equation}
in which the soft-core radius $R(\theta)=\sqrt[6]{C_6(\theta)/2\delta}$ depends on the state- and angle-dependent dispersion coefficient $C_6(\theta)$, and the strength $V_0 = \Omega^4/(8\delta^3)$. The detuning $\delta$ and the potential depth $V_0$ can be controlled by the dressing laser~\cite{SM}. When $C_6(\theta)<0$ and $\delta<0$, the dressed interaction is attractive everywhere. Moreover, it is of cylindrical symmetry with respect to the $z$ axis [Fig.~\ref{fig:introduction}(a)], as $C_6(\theta)$ has no dependence on the azimuth angle [as $m=0$ for all $Y_{lm}(\theta)$ of Eq.~(\ref{eq:C6})]. 

Although $V(d,\theta)$ is attractive for all interparticle distance $d$, its Fourier transform, $\tilde{V}(\mathbf{k})$, is attractive when $k(\theta_k)$ is small and becomes positive when $k(\theta_k) > 1.37\pi/R(\theta_k)$ [Fig.~\ref{fig:introduction}(b)]. It reaches the maximum around $k(\theta_k)\approx 5\pi/3R(\theta_k)$ [Fig.~\ref{fig:introduction}(c)]. As we will show later, the positive region turns out to be important in determining the ground state of the Fermi gas. For convenience, we will use $\bar{k}\approx 5\pi/3\bar{R}$ as the characteristic momentum, where $\bar{R}=R(\pi/2)$ is the soft-core radius along the $x$ axis  [Fig.~\ref{fig:introduction}(a)]. $\tilde{V}(k)$ depends on the modulus $|k|$ as well as on the polar angle (in $k$ space) $\theta_k$. This is in contrast to dipolar interactions whose Fourier transform is solely a function of the polar angle~\cite{marinescu_controlling_1998, yamaguchi_density-wave_2010}.

\textit{Modelling the Rydberg-dressed $^6$Li gas.}
Our system is a three-dimensional (3D) homogeneous gas of $N$ dressed $^6$Li atoms, whose Hamiltonian is given by
\begin{equation} \begin{split}
H &= \int d \mathbf{r} \ \hat\psi^\dag(\mathbf{r})\left[-\frac{\hbar^2}{2m}\nabla^2 -\mu\right]\hat\psi^\dag(\mathbf{r}) \\
&\phantom{{}=} + \iint d \mathbf{r} d \mathbf{r}' \ \hat\psi^\dag(\mathbf{r})\hat\psi^\dag(\mathbf{r}') V(d, \theta) \hat\psi(\mathbf{r}')\hat\psi(\mathbf{r}) ,
\end{split} \end{equation}
where the operator $\hat{\psi}(\mathbf{r})$ annihilates a Fermion with mass $m$ at position $\mathbf{r}$. The first line of the above Hamiltonian gives the usual kinetic energy and chemical potential $\mu$. The second line describes the interaction between two Fermions at  positions $\mathbf{r}$ and $\mathbf{r}'$  via the soft-core potential.  Using the plane-wave basis, the Hamiltonian can be re-expressed as,
\begin{equation} 
	\begin{split}
		H =& \int \frac{ d^3\mathbf{k}}{(2\pi)^3} \left( \frac{\hbar^2|\mathbf{k}|^2}{2m} - \mu \right) \hat{a}^\dag_{\mathbf{k}} \hat{a}_{\mathbf{k}} \\
		& + \iiint \frac{ d^3\mathbf{k}}{(2\pi)^3} \frac{ d^3\mathbf{k}'}{(2\pi)^3} \frac{ d^3\mathbf{q}}{(2\pi)^3} \hat{a}^\dag_{\mathbf{k}+\mathbf{q}} \hat{a}^\dag_{\mathbf{k}'} \tilde{V}(\mathbf{q}) \hat{a}_{\mathbf{k}'+\mathbf{q}} \hat{a}_{\mathbf{k}} ,
	\end{split}
\end{equation}
where $\hat{a}_{\mathbf{k}}=\int d^3 \mathbf{r} \hat{\psi}(\mathbf{r}) e^{-i\mathbf{k}\cdot\mathbf{r}}$ ($\hat{a}^{\dagger}_{\mathbf{k}}=\int d^3 \mathbf{r} \hat{\psi}^\dag(\mathbf{r}) e^{i\mathbf{k}\cdot\mathbf{r}}$) is the annihilation (creation) operator of a Fermion with momentum $\mathbf{k}$.

To understand the impact of the anisotropic interaction on the many-body physics, we study the ground state of the system within the Hartree-Fock-Bogoliubov (HFB) approach~\cite{zhao_hartree-fock-bogoliubov_2010-1,ring_nuclear_1980}. Assuming pairing occurs between particles with momentum $\mathbf{k}$ and $-\mathbf{k}$, this allows to write down the BCS wave function $|G\rangle_{\rm BCS} = \prod_{\mathbf{k}} \left( u_\mathbf{k} + v_\mathbf{k} \hat{a}^\dag_{\mathbf{k}} \hat{a}^\dag_{-\mathbf{k}} \right) |0\rangle$. The ground state is determined by the approximate HFB Hamiltonian,
\begin{equation}
	H_{\rm HFB} =  \int \frac{ d^3\mathbf{k}}{(2\pi)^3} \  \xi_\mathbf{k} \hat{a}^\dag_{\mathbf{k}}\hat{a}_{\mathbf{k}} + \left[\Delta(\mathbf{k}) \hat{a}^\dag_{\mathbf{k}}\hat{a}^\dag_{-\mathbf{k}} + \text{H.c.} \right],
	\label{eq:HBF}
\end{equation}
where $\xi_\mathbf{k}=\varepsilon_k + U_e(\mathbf{k}) +U_d-\mu$ with $\varepsilon_k=\hbar^2|\mathbf{k}|^2/{2m}$ being the kinetic energy of free fermions. The Hartree energy $U_d=N\tilde{V}(0)$ is a constant and can be absorbed into the chemical potential. Both the Fock energy $U_e(\mathbf{k}) = - \sum_{\mathbf{k}'} \tilde{V}(\mathbf{k}-\mathbf{k}')\langle \hat{a}^\dag_{\mathbf{k}'}\hat{a}_{\mathbf{k}'} \rangle$ and the gap function $	\Delta(\mathbf{k}) = \sum_{\mathbf{k}'} \tilde{V}(\mathbf{k}-\mathbf{k}') \langle \hat{a}_{-\mathbf{k'}}\hat{a}_{\mathbf{k}'} \rangle$ depend on $\tilde{V}(\mathbf{k})$. The shape of the interaction, i.e., $\tilde{V}(-\mathbf{k})=\tilde{V}(\mathbf{k})$, implies $U_e(-\mathbf{k})=U_e(\mathbf{k})$  and $\Delta(-\mathbf{k})=-\Delta(\mathbf{k})$. In the absence of two-body interaction, $\tilde{V}(\mathbf{k})=0$, one obtains a spherical FS of radius $k_F=(6\pi^2 \rho)^{1/3}$  ($\rho$ is the real space density) in momentum space, where the density distribution is $\tilde{\rho}_0(\mathbf{k})=\Theta(|{k}_F-\mathbf{k}|)$ in the ground-state with $\Theta(\cdot)$ to be the Heaviside function. As the variational parameter $u_{\mathbf{k}}$ and $v_{\mathbf{k}}$ satisfy
\begin{gather}
	u_\mathbf{k}^2 = \frac12 \left[ 1 + \frac{\xi_\mathbf{k}}{\sqrt{\xi_\mathbf{k}^2 + |\Delta_\mathbf{k}|^2}} \right],\,
	v_\mathbf{k}^2 = \frac12 \left[ 1 - \frac{\xi_\mathbf{k}}{\sqrt{\xi_\mathbf{k}^2 + |\Delta_\mathbf{k}|^2}} \right],\nonumber
\end{gather}  
 we can obtain the BCS ground state by solving the gap function self-consistently. 

\begin{figure}
	\centering
	\includegraphics[width=\linewidth]{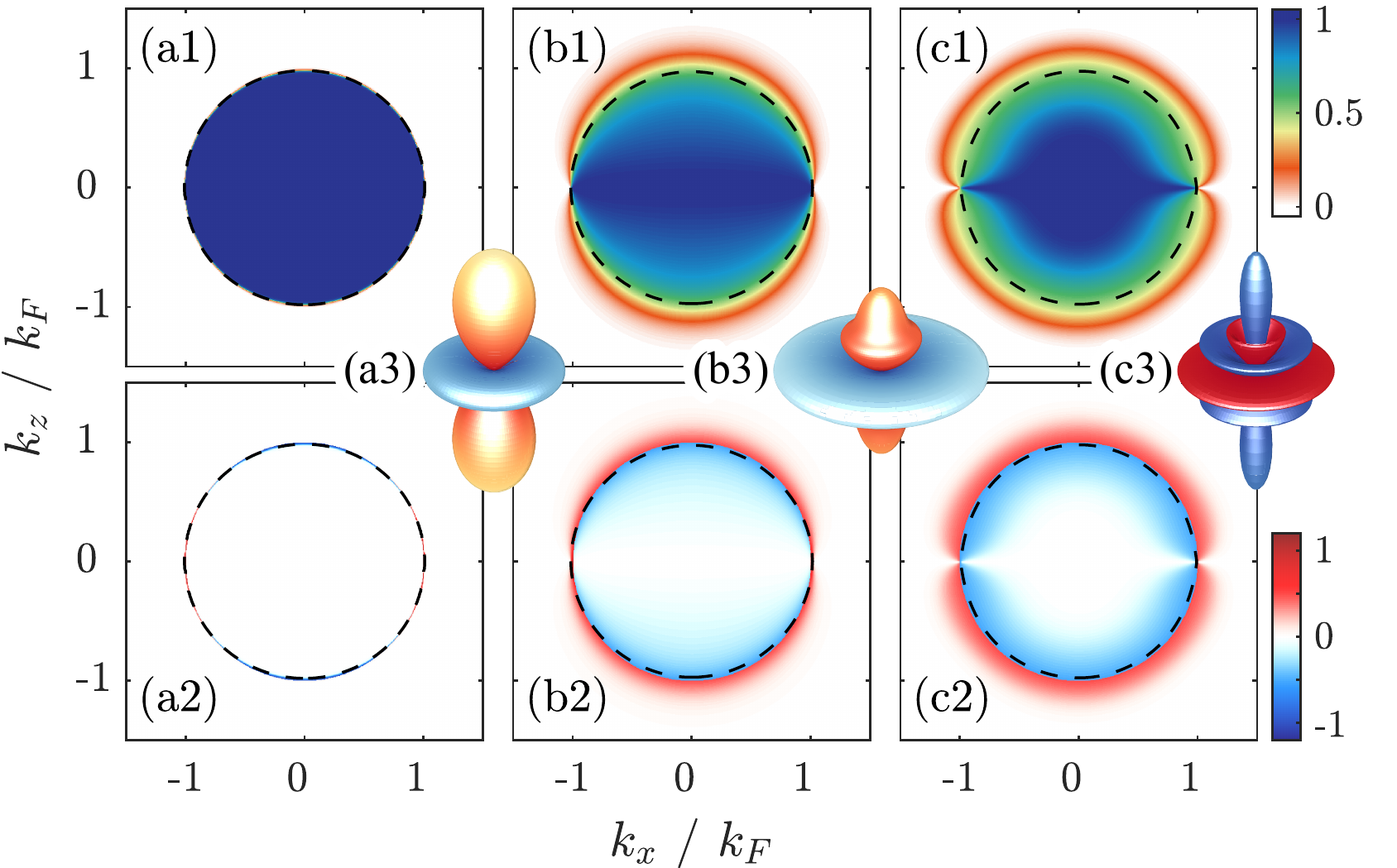}
	\caption{ State-dependent FS deformation. Density distribution in momentum space for (a1)  $n=25$ with $\bar{R} = 1.12~{\rm \mu m}$, (b1) $n=35$ with $\bar{R} = 2.08~{\rm \mu m}$, and (c1) $n=55$ with $\bar{R} = 4.77~{\rm \mu m}$. The corresponding $\delta\tilde{\rho}(\mathbf{k})$ is given in panels (a2), (b2), and (c2). The dashed line in the first and second row is the fitted FS (see \textbf{SM}~\cite{SM} for details), which locates the momentum where $\delta\tilde{\rho}(\mathbf{k})\approx 0$. For $n=25$, the deformation of the FS is described by the spherical harmonics $Y_{20}(\theta_k)$, shown in panel (a3). For $n=35$ and $55$, the FS deformation has complicated multipolar features, as shown in panels (b3) and (c3), respectively. Here $k_F=2~\mu$m$^{-1}$, and other parameters are same with Fig.~\ref{fig:introduction}.  }
	\label{fig:example}
\end{figure}

\textit{Multipolar FS deformation.} 
We first illustrate with examples that the anisotropic Rydberg-dressed interaction can break the spherical symmetry of the FSs for noninteracting fermions. In Fig.~\ref{fig:example}(a1), we show the momentum distribution for $n=25$ (i.e., the atoms are dressed to the $|25D_{5/2}, 5/2\rangle$ state), which displays a sharp edge at the FS. The FS is slightly deformed from a sphere, which is confirmed by a numerical fit of the FS based on the conventional mean-field solution (see the Supplemental Material, \textbf{SM}, for details~\cite{SM}). This anisotropy results from the fact that the scattering of two atoms does not preserve each of their orbital angular momenta due to the anisotropic dressed interaction~\cite{you_prospects_1999}. However the difference of the momentum distribution from the interaction-free one, given by $\delta \tilde{\rho}(\mathbf{k})=\tilde{\rho}(\mathbf{k})-\tilde{\rho}_0(\mathbf{k})$, is marginal for the low-lying $n=25$ state [Fig.~\ref{fig:example}(a2)]. 

The momentum distribution changes qualitatively with increasing principal quantum number $n$. For $n=35$ [Fig.~\ref{fig:example}(b1)] and $55$ [Fig.~\ref{fig:example}(c1)], the Fermi sea is depleted notably around $k_F$. Importantly, the momentum distributions become strongly anisotropic.
To quantify the anisotropy of $\tilde{\rho}(\mathbf{k})$, we first evaluate its angular distribution by integrating over the radial part, $\chi(\theta_k)=2\pi/N\int \tilde{\rho}(\mathbf{k})k^2dk$ with the normalization condition $\int_{0}^{\pi} \chi(\theta_k)\sin\theta_k d\theta_k=1$. Here, the azimuthal angle $\phi_k$ has been integrated out straightly due to the cylindrical symmetry of $\tilde{\rho}(\mathbf{k})$. Then the angular function is expanded in terms of spherical harmonics, $\chi(\theta_k)=\sum_{l=0,2,4...} \lambda_l Y_{l0}(\theta_k)$ where $\lambda_l=\int \chi(\theta_k)Y_{l0}(\theta_k)\sin\theta_k d\theta_k$~\cite{SM} is the coefficient of $\chi(\theta_k)$ projected to $Y_{l0}(\theta_k)$. Here, a nonvanishing $\lambda_l$ ($l>0$) means that the shape of the FS is not spherical. For parameters considered in this work, it is found that $\lambda_l$ becomes negligible when $l>6$. 

\begin{figure}
	\centering
	\includegraphics[width=0.95\linewidth]{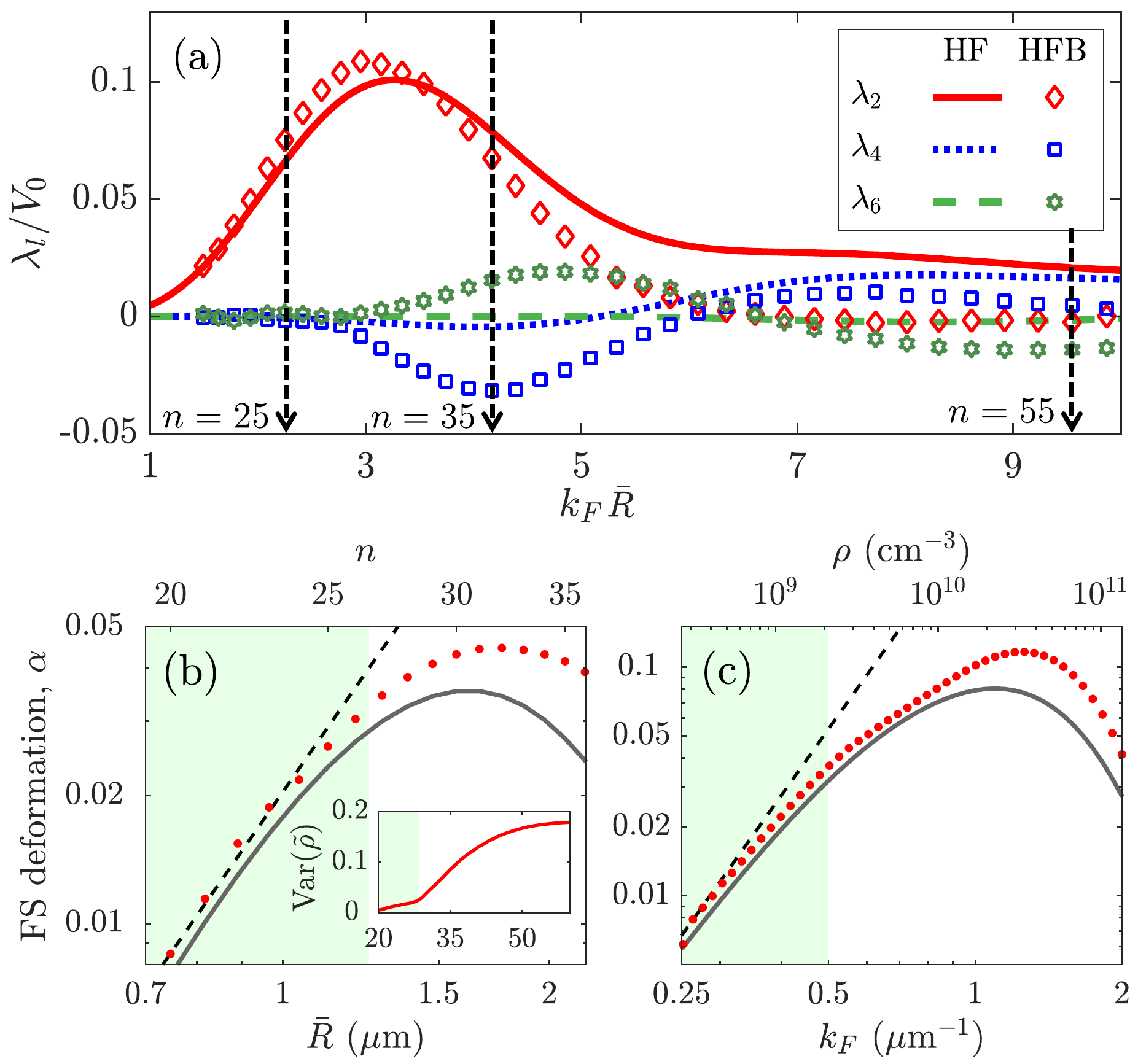}
	\caption{ Characterization of the multipolar FS.  (a) The projection coefficient $\lambda_j$ (\textcolor{red}{diamond}, \textcolor{blue}{square}, \textcolor{ForestGreen}{star}) and $\lambda_j'$ (\textcolor{red}{solid}, \textcolor{blue}{dotted},  \textcolor{ForestGreen}{dashed}) based on the HFB and HF calculation. In the calculation, we vary $\bar{R}$ (through $n$) and fix $k_F=2~{\rm \mu m}^{-1}$.  $\lambda_2$ dominates when $k_F\bar{R}< 3$ ($n<30$). $\lambda_4$ and $\lambda_6$ become important when $k_F\bar{R}>3$, where the multipolar FS deformation emerges.  When $k_F\bar{R} \lesssim 3$, the deformation of the FS can also be characterized by its ellipticity, $\alpha = 1-\mathbf{k}_F(0)/\mathbf{k}_F(\pi/2)$.  In panels (b) and (c), we show the ellipticity by changing the soft-core radius $\bar{R}$ ($n$) and $k_F$ ($\rho$), respectively.  When $\lambda_2$ dominates, one finds $\alpha\approx 3\sqrt{5/\pi}\lambda_2/4$, establishing a relation of the ellipticity and $\lambda_2$.  The numerical data (dots) and perturbation results (solid) follow the power-law scaling well in the shaded area.  The inset in panel (b) shows the variance of the density distribution $\tilde{\rho}(\mathbf{k})$. In (b) $k_F=2~{\rm \mu m}^{-1}$ and (c) $n=35$. Other parameters are same as in Fig.~\ref{fig:introduction}. } 
	\label{fig:ellipticity}
\end{figure}

The main result of this work is that the shape of the FS involves multiple $\lambda_l$. As shown in Fig.~\ref{fig:ellipticity}(a), $\lambda_2$ is the only nonzero projection coefficient when $k_F\bar{R}<3$ ($k_F\lesssim  0.6\bar{k}$) and assumes its maximal value around $k_F\bar{R}\approx 3$. In this region, we obtain a \textit{dipolar deformation} ($l=2$), where the FS is stretched in the $k_z$ direction [Fig.~\ref{fig:example}(a3)], and becomes an ellipsoid.  When $k_F\bar{R}>3$, the amplitudes of $\lambda_4$ and $\lambda_6$ increase while the one for $\lambda_2$ decreases. In the region $3.5\lesssim k_F\bar{R} \lesssim 5$,  $|\lambda_4|$ and $\lambda_6$ become comparable to $\lambda_2$. In this \textit{multipolar deformation} regime, mainly three spherical harmonics contribute to the FS: $Y_{20}(\theta_k)$, $Y_{40}(\theta_k)$, and $Y_{60}(\theta_k)$ [Fig.~\ref{fig:example}(b3)]. When further increasing $k_F\bar{R}>6$, the value of all $\lambda_l$ decreases and saturates. As $|\lambda_4|>|\lambda_6|>|\lambda_2|$, this yields a FS deformation significantly different [Fig.~\ref{fig:example}(c3)] from the one shown in Fig.~\ref{fig:example}(a3).

The depletion of the momentum density becomes significant when the multipolar deformation is present. This leads to an increase of the variance ${\rm Var}(\tilde{\rho})=  \sum_{\mathbf{k}}[\tilde{\rho}(\mathbf{k})-\tilde{\rho}_0(\mathbf{k})]^2/\rho$ of the momentum distribution, as shown in the inset of Fig.~\ref{fig:ellipticity}(b). However, when $n<30$, the variance is small and increases only slowly with $n$. When $n>30$, it increases rapidly and saturates gradually, where the deformation is given mainly by the $\lambda_6$ term [Fig.~\ref{fig:ellipticity}(a)].

\begin{figure}
	\centering
	\includegraphics[width=0.98\linewidth]{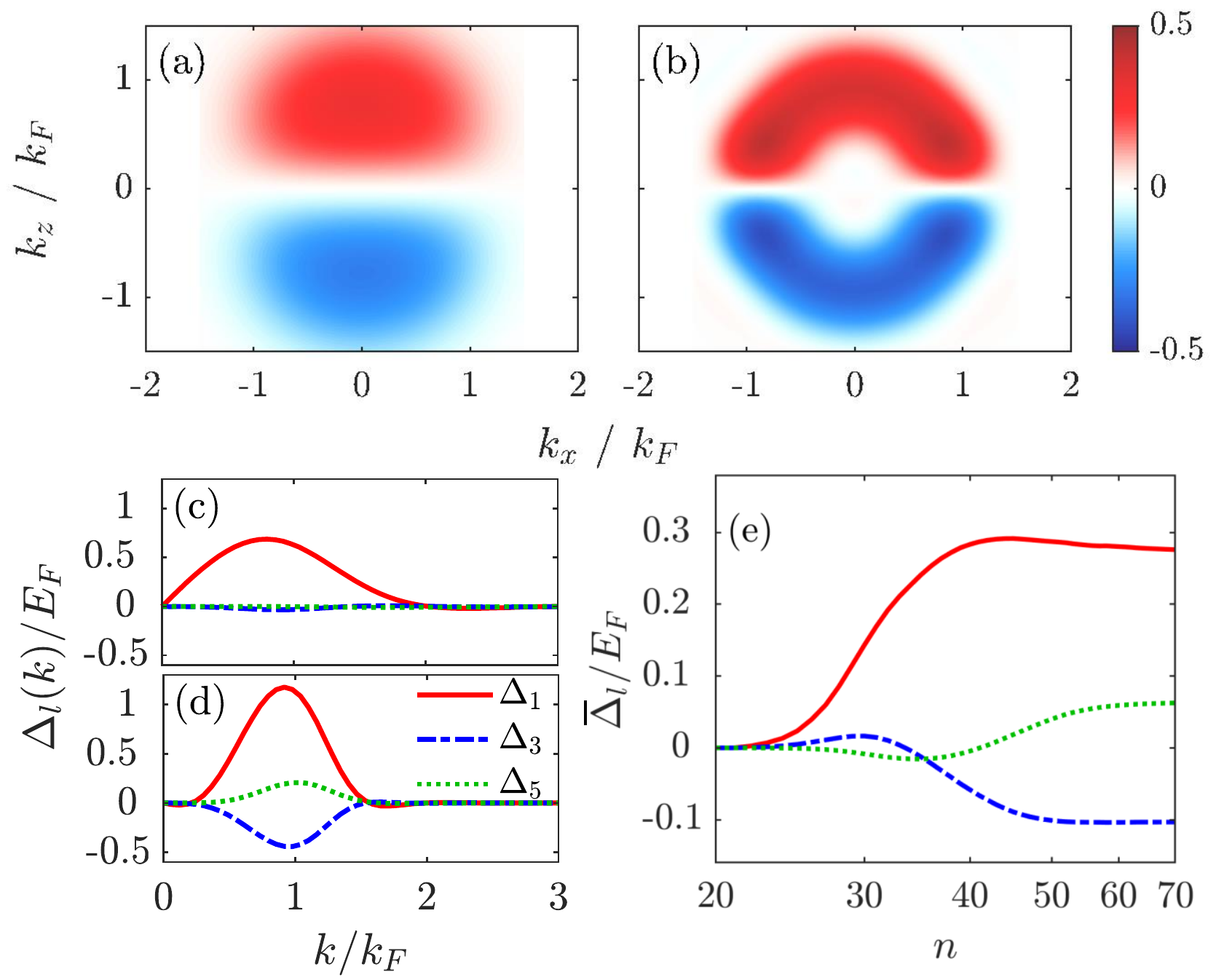}
	\caption{ Cooper pairs of multiple partial waves. Gap functions $\Delta(\mathbf{k})$ for (a) $n=35$ and (b) $n=55$. The projection of $\Delta(\mathbf{k})$ to the spherical harmonics $Y_{l0}$ ($l$ odd) for (b) $n=35$ and (c) $n=55$, respectively. For panel (c), $n=35$, the gap function is occupied by the $p$-wave component. For panel (d), $n=55$, it turns out that $f$-wave ($l=3$) and $h$-wave ($l=5$) pairing become important as well. (e) The population of various pair states as a function of quantum number $n$.  Note that $f$- and $h$-wave pairing are non-negligible when $n>40$. Other parameters are the same as those in Figs.~\ref{fig:ellipticity}(b) and \ref{fig:ellipticity}(c), respectively.  }
	\label{fig:pairs}
\end{figure}

\textit{Scaling in the dipolar deformation regime.}
The value of $\lambda_2$, which parametrizes the FS deformation, follows a power-law scaling with respect to the system parameters, such as the interaction strength $V_0$, the density of atoms $\rho$, and the interaction length $\bar{R}$. When $\lambda_2$ dominates, we find that the density distribution changes rapidly at the FS which gives rise to a modified Fermi momentum $\mathbf{k}'_F$ [Fig.~\ref{fig:example}(a1)]. This deformation is quantified by the deviation of $\mathbf{k}'_F$ from $k_F$, $\delta \mathbf{k}_F= \mathbf{k}'_F-{k}_F\vec{\mathbf{n}}$, where  $\vec{\mathbf{n}}$ is the unit vector parallel to $\mathbf{k}'_F$. As pairing is not important here, $\delta \mathbf{k}_F$ can be evaluated from the Hartree-Fock  energy~\cite{Chan2010, Ronen2010} 
\begin{equation} \label{eq:Perturbation}
	\frac{\delta \mathbf{k}_F}{k_{F}} \approx  \frac{1}{2E_{F}} \int d^3\mathbf{q} \tilde{V}(\mathbf{k}'_{F}-\mathbf{q}) \tilde{\rho}(\mathbf{q}).
\end{equation}
Assuming $\tilde{\rho}(\mathbf{q})\approx \tilde{\rho}_0(\mathbf{q})$ and projecting $\delta \mathbf{k}_F$ onto spherical harmonics, this yields
\begin{equation}
	\label{eq:projFS}
	\frac{\delta \mathbf{k}_F}{k_F } \approx  \frac{4\pi}{3} \sum_{l=2,4,6} \lambda_l Y_{l0}(\theta_k) ,
\end{equation}
with $\lambda_l = 3/(8\pi E_F) \int_0^\infty F_l\left({k}/{k_F \bar{R}}\right) G_l\left(k\right) dk$ being the approximate projection coefficient. Here, we have defined $F_l(k) =  \int_0^\pi Y_{l0}(\theta_k) V(k\bar{R},\theta_k)\sin\theta_k{\rm d}\theta_k $ and  $G_l(k) = 4 {\rm i}^l k j_1(k) j_l(k)$, where $j_l(k)$ is the spherical Bessel function of the first kind (see the \textbf{SM}~\cite{SM}).

As shown in Fig.~\ref{fig:ellipticity}(a), the coefficient $\lambda_2$, obtained from the perturbative calculation agrees well with the full-numerical method up to $k_F\bar{R} \approx 3$. 
In this region, we can make a Taylor expansion of $\lambda_2$ in terms of $k_F\bar{R}$, whose leading term is $\lambda_2 \propto V_0  (k_F \bar{R})^5/E_F$~\cite{SM}. This results in the power law relations $\lambda_2 \propto \bar{R}^3, n^{11/2}, k_F ^{3}$. This scaling matches with the numerical calculations when $k_F\bar{R}$ is small, as shown by the shaded area in Figs.~\ref{fig:ellipticity}(b) and \ref{fig:ellipticity}(c). Experimentally, the predicted scaling can be measured by tuning the atomic density ($k_F$) and laser parameters ($n$ and $\delta$) jointly or separately, which is useful in exploring the FS deformation and the dressed interaction experimentally. Finally, we note that the peak at $k_F\bar{R}\approx 3$ can also be explained within the perturbative treatment pursued here: $F_2\left[{k}/(k_F  \bar{R})\right]$ has a single peak located at $k=k_F\bar{R}$, and $G_2(k)$ oscillates with $k$, as discussed in the \textbf{SM}~\cite{SM}. We thus obtain the maximal $\lambda_2$ when the peaks of $F_2(k)$ and $G_2(k)$ coincide at $k_F\bar{R}\approx 3$.

\textit{Cooper pairing in the multipolar deformation regime.}
In the HFB approach, the gap function $\Delta(\mathbf{k})$ serves as an order parameter of superfluidity. Here the superfluid is formed by Cooper pairs of two fermionic atoms with opposite momenta~\cite{you_prospects_1999, zhao_hartree-fock-bogoliubov_2010-1}. In Fig.~\ref{fig:pairs}(a), we show the distribution of the gap function $\Delta(\mathbf{k})$ for $n=35$. It is nonzero in a wide region around the FS where the momentum density differs drastically from the interaction-free Fermi sea [Fig.~\ref{fig:example}(b1)]. Here an antisymmetry of the gap function is observed along the $k_z$ axis. For $n=55$ [Fig.~\ref{fig:pairs}(b)], the pair distribution becomes more confined around the FS, and its peak value increases [Fig.~\ref{fig:example}(c1)].

To gain further insights of the pairing, we expand the gap function into partial waves, $\Delta(\mathbf{k})=\sum_l\Delta_{l}(k) Y_{lm}(\theta_k)$
with $\Delta_{l}(k) = \int \Delta(\mathbf{k}) Y_{lm}(\theta_k, \phi_\mathbf{k}) d \Omega_{\mathbf{k}}$ being the $l$-wave pair state ($l$ is an odd integer)~\cite{you_prospects_1999}. As shown in Fig.~\ref{fig:pairs}(c), $p$-wave pairing is dominant when $n=35$.  Importantly, new pairing states emerge when the FS is deformed strongly. As depicted in Fig.~\ref{fig:pairs}(d), $f$-wave ($l=3$) and $h$-wave ($l=5$) pairing, together with $p$-wave pairing, are found when $n=55$. A systematic study shows that the population of the Cooper pairs can be enhanced by dressing to higher (more strongly interacting) Rydberg states.  In Fig.~\ref{fig:pairs}(e), the total  population of individual pair states, $\bar{\Delta}_l=\int{d^3\mathbf{k}} \Delta_l(\mathbf{k})$, is shown. When $n$ increases, the occupation of $p$-, $f$-, and $h$-wave pairs also increases. In particular, considerable $f$- and $h$-wave populations are obtained when $n>40$.

\textit{Discussion and outlook.}
The quantum simulation of $p$- and $f$-wave pairing with ultracold atoms has attracted much attention due to their importance in understanding superfluidity in $^3$He-$A$~\cite{israelsson_f-wave_1986}, $^3$He-$B$ phase~\cite{davis_discovery_2008}, and unconventional superconductors~\cite{stewart_unconventional_2017}. It has been shown that $p$-wave pairing can be realized with ultracold fermions with the dipolar interaction~\cite{ho2005, Cheng2006, zhao_hartree-fock-bogoliubov_2010-1, PhysRevLett.103.155302, fedorov_novel_2016-1} or isotropic dressed interaction in three dimensions~\cite{xiong_topological_2014-1}. However, $f$-wave pairing can only occur in 2D under extra restricted conditions, including Bose-Fermi mixed interactions~\cite{lee_f-wave_2010,mathey_exotic_2007}, excited bands~\cite{hung_frustrated_2011}, and repulsive Rydberg-dressed interactions~\cite{PhysRevA.96.053611,keles_f-wave_2020}. Our results show that an anisotropic Rydberg-dressed interaction allows to achieve such unconventional pairing states in three dimensions (3D). Moreover, it even provides access to the $h$-wave pairing channel.

Such $h$-wave pairing has not been thoroughly studied and is certainly worth further exploration. The anisotropic and attractive dressed interaction opens further opportunities to probe novel phases, such as the nematic phase~\cite{kivelson_electronic_1998}, topological superfluid~\cite{PhysRevLett.103.155302, PhysRevA.86.031603, liu_weyl_2015}, and supersolid phases~\cite{PhysRevA.85.021604, PhysRevB.89.174511}, as well as to probe nonequilibrium dynamics driven by long-range anisotropic interactions~\cite{PhysRevLett.113.210402, guardado-sanchez_quench_2020}. Finally, we remark that anisotropic but repulsive interactions~\cite{SM} can be obtained in the Rydberg dressing of $^{39}$K atoms~\cite{arias_realization_2019}. Such repulsive interaction allows for probing the Kohn-Luttinger mechanism~\cite{shechtman_metallic_1984, maiti_superconductivity_2013, cao_kohn-luttinger_2020}.   

\textit{Acknowledgments.}
We are thankful for useful discussions with Yongqiang Li, Vijay Shenoy, S. Kumar Mallavarapu, and Gary McCormack. Y.Z. acknowledges support from National Natural Science Foundation of China (Grant No. 12088101) and NSAF (Grant No. U1930403). W.L. acknowledges support from the EPSRC through Grant No. EP/R04340X/1 via the QuantERA project ``ERyQSenS'', the UKIERI-UGC Thematic Partnership (IND/CONT/G/16-17/73), and the Royal Society through the International Exchanges Cost Share Award No. IEC$\backslash$NSFC$\backslash$181078. R.N.  acknowledges DST-SERB for Swarnajayanti Fellowship File No. SB/SJF/2020-21/19.  I.L. acknowledges support from the ``Wissenschaftler R\"{u}ckkehrprogramm GSO/CZS'' of the Carl-Zeiss-Stiftung and the German Scholars Organization e.V., as well as the Deutsche Forschungsgemeinschaft through SPP 1929 (GiRyd), Grant No. 428276754, and the ``Internationale Spitzenforschung'' program of the BW Foundation. We are grateful for access to the Augusta High Performance Computing Facility at the University of Nottingham.

%\bibliography{references}


%%%%%%%%%% Merge with supplemental materials %%%%%%%%%%
\onecolumngrid
\clearpage
\widetext
\begin{center}
	\textbf{\large Supplementary material for:\\ Multipolar Fermi-surface deformation in a Rydberg-dressed Fermi gas with long-range anisotropic interactions}\\
	\vspace*{0.5cm}
	Yijia Zhou$^{1,2}$, Rejish Nath$^3$, Haibin Wu$^{4,5}$, Igor Lesanovsky$^{1,6}$, Weibin Li$^1$\\
	\vspace{0.2cm}
	$^1$\textit{School of Physics and Astronomy and Centre for the Mathematics and \\ Theoretical Physics of Quantum Non-equilibrium Systems, University of Nottingham, Nottingham, NG7 2RD, UK}\\
	$^2$\textit{Graduate School of China Academy of Engineering Physics, Beijing, 100193, China}\\
	$^3$\textit{Department of Physics, Indian Institute of Science Education and Research, \\ Dr. Homi Bhabha Road, Pune- 411008, Maharashtra, India}\\
	$^4$\textit{State Key Laboratory of Precision Spectroscopy, East China Normal University, Shanghai 200062, China}\\
	$^5$\textit{Collaborative Innovation Center of Extreme Optics, Shanxi University, Taiyuan 030006, China}\\
	$^6$\textit{Institut f\"ur Theoretische Physik, University of T\"ubingen, 72076 T\"ubingen, Germany}
\end{center}
%%%%%%%%%% Merge with supplemental materials %%%%%%%%%%
%%%%%%%%%% Prefix a "S" to all equations, figures, tables and reset the counter %%%%%%%%%%
\setcounter{equation}{0}
\setcounter{figure}{0}
\setcounter{table}{0}
\setcounter{page}{1}
\makeatletter
\renewcommand{\theequation}{S\arabic{equation}}
\renewcommand{\thefigure}{S\arabic{figure}}
\renewcommand{\bibnumfmt}[1]{[S#1]}
\renewcommand{\citenumfont}[1]{S#1}
%%%%%%%%%% Prefix a "S" to all equations, figures, tables and reset the counter %%%%%%%%%%
\begin{center}
	\begin{minipage}{0.8\textwidth} 
		\quad This supplementary material gives details on the analysis in the main text and provides additional data for calculating the dressed interaction potential.
	\end{minipage}
\end{center}

\begin{figure} [!ht]
	\centering
	\includegraphics[width=0.9\linewidth]{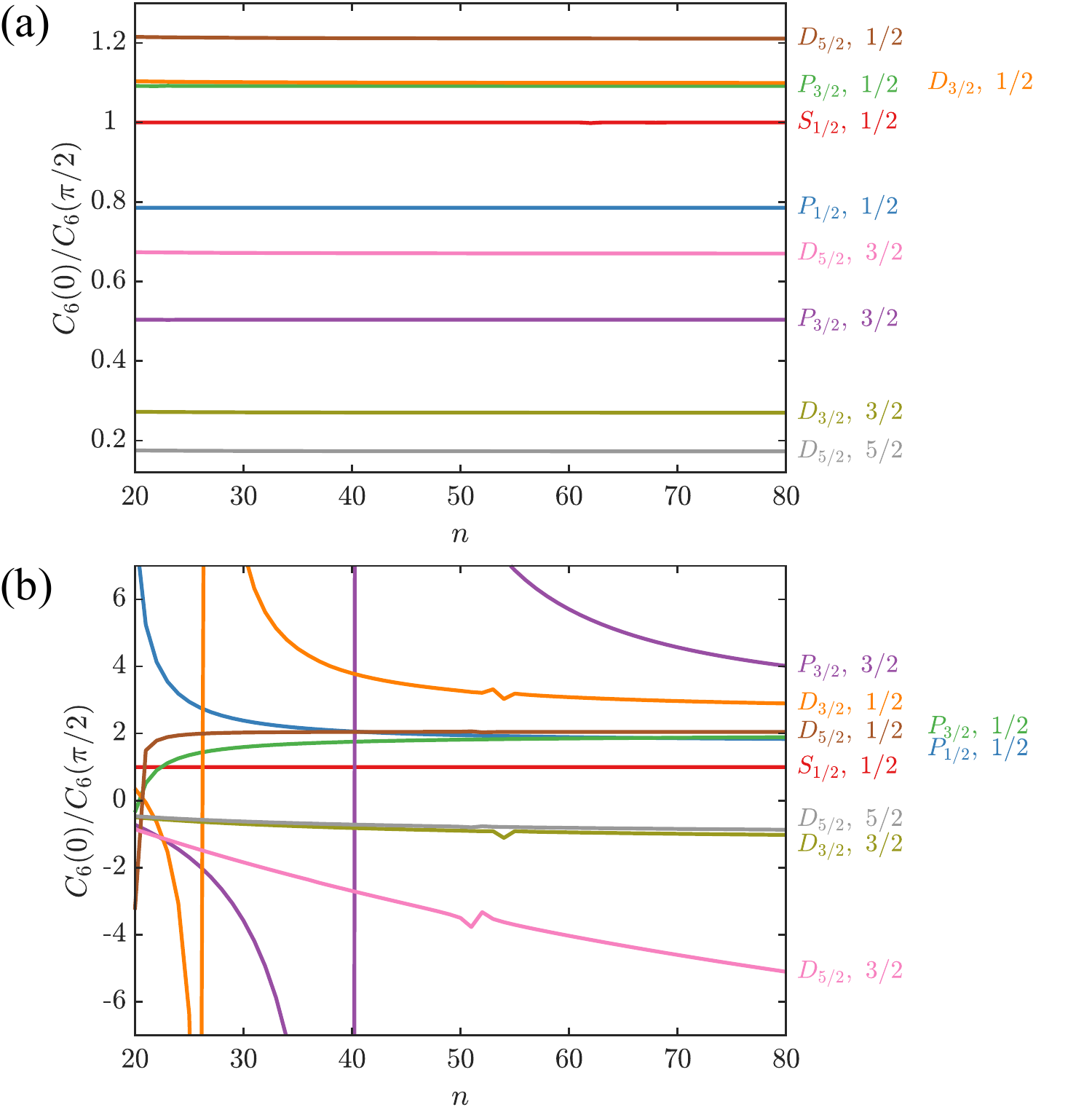}
	\caption{ Ratio ${C_6}(0)/{C_6}(\pi/2)$ for various Rydberg states of (a) $^6$Li and (b) $^{40}$K atoms. (a) For $^6$Li, the van der Waals interaction is the most anisotropic in $|nD_{5/2},m_j=5/2\rangle$ state, and its anisotropy ${C_6}(0)/{C_6}(\pi/2) \approx 0.173$  is largely insensitive to the principal quantum number $n$. (b) For $^{40}$K, the sign of $C_6(\pi/2)$ flips at $n=40\rightarrow41$ for $|P_{3/2},m_j=3/2\rangle$, and $n=26\rightarrow27$ for $|D_{3/2},m_j=1/2\rangle$. This means that the ratio near the resonant states is particularly large. For example, At $|41P_{3/2},m_j=3/2\rangle$, ${C_6}(0)/{C_6}(\pi/2) \approx 261$, and At $|27D_{3/2},m_j=1/2\rangle$, ${C_6}(0)/{C_6}(\pi/2) \approx 63$. }
	\label{fig:C6ratios}
\end{figure}

\section{Dispersion coefficients of the Rydberg states}
The dispersion coefficient ${C_6(\theta)}$ is calculated with the \textsf{Alkali-Rydberg-Calculator (ARC)} package~\cite{s_Sibalic2017, s_Robertson2021}. In the Rydberg $|nD_{5/2}\rangle$ state, the van der Waals interaction is cylindrically symmetric, i.e.   ${C_6}(\theta)$ is dependent on the polar angle $\theta$ but not on the azimuth angle $\phi$.  In the Rydberg-dressing, one would like to gain strong soft-core interactions while maintaining a small excitation in the Rydberg state, whose population is approximately given by $\Omega^2/4\delta^2$.  In a recent Rydberg dressing experiment with $^6$Li atoms, it was shown that relatively strong interactions can be achieved with $\Omega\lesssim0.3\delta$~\cite{s_guardado-sanchez_quench_2020}. In our simulation, we consider parameters $\Omega\approx 0.15\delta$ throughout.  
Furthermore,  coefficient $C_6(\theta)$ should have same sign for all $\theta$ in order to avoid resonances~\cite{s_lan_emergent_2015, s_lan_devils_2018-1}.  Our calculations show that this is always guaranteed for $^6$Li.
\begin{figure} [ht]
	\centering
	\includegraphics[width=0.6\linewidth]{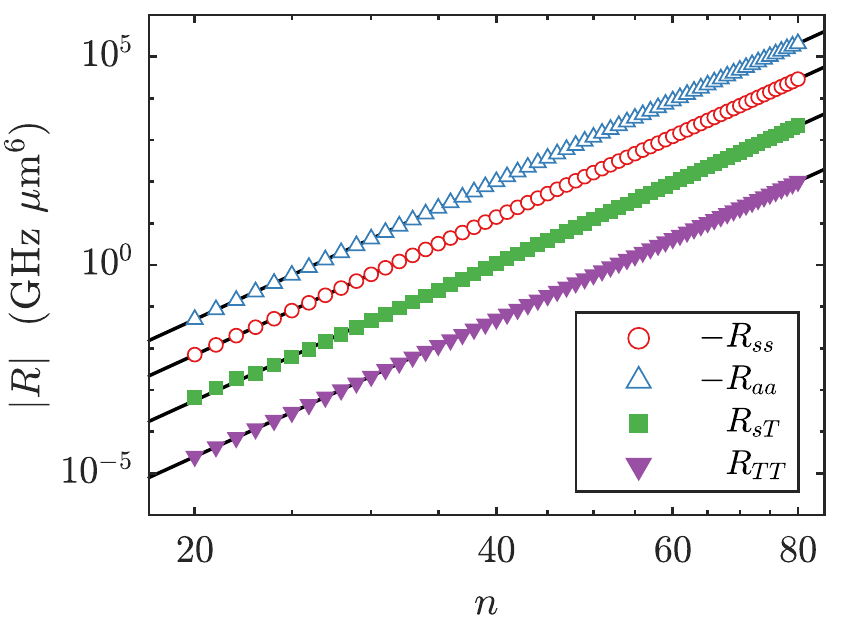} 
	\caption{(a) Irreducible components for ${C_6}$ parameters of $^6$Li at $nD_{5/2}$ states. $R_{ss}$ is for the scalar component, $R_{aa}$ is for the axial component, $R_{sT}$ is for the scalar-tensor component, and $R_{TT}$ is for the tensor-tensor component. Each component is fitted by $R\sim n^{11}$ (solid lines). }. 
	\label{fig:Rn_Li6}
\end{figure}

The anisotropy of the Rydberg atom interaction depends on the angular momentum quantum number of the Rydberg state. In $nS$ states, the anisotropy is typically negligible. In order to find the maximal anisotropic interacting states, we have examined both the $nP$ and $nD$ states. The ratio of ${C_6}(\theta)$ at $\theta=0$ and $\pi/2$ is used to quantify the strength of the anisotropy. Strong anisotropic interactions will be obtained when the ratio is much larger or smaller than 1. As shown in Fig.~\ref{fig:C6ratios}, we found that the most anisotropic interaction appears in $nD_{5/2}$ states with $m_j=5/2$, where ${C_6}(0)/{C_6}(\pi/2) \approx 0.17$. Therefore we will focus on  $|nD_{5/2},m_j=5/2\rangle$ states of $^6$Li atoms in the following discussion. Another important feature here is that this ratio is not sensitive to the principal quantum number $n$ for $^6$Li atoms. In contrast, the anisotropy of $^{40}$K atoms varies strongly when varying $n$, see Fig.~\ref{fig:C6ratios}(b). In $|nP_{3/2}\rangle$ states, we find repulsive interactions between Rydberg atoms, i.e. $C(\theta)>0$. This state can be used to create anisotropic and repulsive dressed interactions with $^{40}$K atoms. 

The angular dependent coefficient ${C_6}(\theta)$ can be expressed in the following form~\cite{s_kamenski_energy_2017},
\begin{eqnarray}
	{C_6}(\theta) &=& R_{ss} - \frac{m_j^2(3\cos^2\theta-2)}{12J^2}R_{aa} +  \frac{3m_j^2-J(J+1)(3\cos^2\theta-1)}{2J(2J-1)}R_{sT}\nonumber \\
	& +&  \frac{3}{2}\left[\frac{3m_j^2-J(J+1)}{2J(2J-1)}\right]^2(9\cos^4\theta-8\cos^2\theta _1)R_{TT},
\end{eqnarray}
where the coefficients $R_{ss}, R_{aa}$, $R_{sT}$ and $R_{TT}$ are irreducible components of the van der Waals coefficient, and are called scalar, axial, scalar-tensor and tensor-tensor components, respectively. Their numerical values are shown in Fig.~\ref{fig:Rn_Li6}. These data can be fitted with the power-law form, given by $[-3.4824,\ -24.1858,\ 0.3262,\ 0.0135]\times10^{-17}n^{11}~{\rm GHz~\mu m^6}$.

Due to the cylindrical symmetry of the van der Waals interaction, we can express $C(\theta)$ in terms of spherical harmonics $Y_{l0}(\theta,\phi)$, 
\begin{equation}
	C_6(\theta)=\mathcal{C}_0Y_{00}(\theta,\phi)+\mathcal{C}_2Y_{20}
	(\theta,\phi) + \mathcal{C}_4 Y_{40}(\theta,\phi),
\end{equation}
where we have defined parameters $C_j$ ($j=0,2,4$),
\begin{equation}
	\begin{split}
		\mathcal{C}_0 &= 2\sqrt{\pi} R_{ss} + \frac{m_j^2}{6J^2}\sqrt{\pi}R_{aa} +\frac{3m_j^2}{J(2J-1)}\sqrt{\pi}R_{sT}-\frac{13}{5}\left[\frac{3m_j^2-J(J+1)}{2J(2J-1)}\right]^2\sqrt{\pi}R_{TT} , \\
		\mathcal{C}_2 &=-\sqrt{\frac{\pi}{5}}\frac{m_j^2}{3J^2}R_{aa} - \sqrt{\frac{\pi}{5}}\frac{2J+2}{2J-1}R_{sT} + \sqrt{\frac{\pi}{5}} \frac{4}{9}\left[\frac{3m_j^2-J(J+1)}{2J(2J-1)}\right]^2R_{TT} ,  \\
		\mathcal{C}_4 &= \frac{72\sqrt{\pi}}{35}\left[\frac{3m_j^2-J(J+1)}{2J(2J-1)}\right]^2R_{TT}.
	\end{split}
\end{equation}
In the above expression $m_j=0$ due to the cylindrical symmetry, which means there is no dependence on azimuth angle $\phi$ in the spherical harmonics. For the sake of simplicity, we rewrite $Y_{l0}(\theta,\phi)\equiv Y_{l0}(\theta)$.

\section{Details on the Hartree-Fock-Bogoliubov approximation}
The Hamiltonian in momentum space
\begin{equation} \label{eq:Hamiltonian}
	H = \int \frac{{\rm d}^3\mathbf{k}}{(2\pi)^3} \frac{|\mathbf{k}|^2}{2m} \hat{a}^\dag_{\mathbf{k}}\hat{a}_{\mathbf{k}} + 
	\iiint \frac{{\rm d}^3\mathbf{k}}{(2\pi)^3} \frac{{\rm d}^3\mathbf{k}'}{(2\pi)^3} \frac{{\rm d}^3\mathbf{q}}{(2\pi)^3} \hat{a}^\dag_{\mathbf{k}+\mathbf{q}} \hat{a}^\dag_{\mathbf{k}'} \tilde{V}(\mathbf{q}) \hat{a}_{\mathbf{k}'+\mathbf{q}}\hat{a}_{\mathbf{k}},
\end{equation}
where $\tilde{V}(\mathbf{q})$ is the Fourier transform of the two-body interaction potential $V(\mathbf{r})$. According to the mean-field theory, the four-operator term is reduced to
\begin{align*}
	\hat{a}^\dag_{\mathbf{k}+\mathbf{q}} \hat{a}^\dag_{\mathbf{k}'} \hat{a}_{\mathbf{k}'+\mathbf{q}}\hat{a}_{\mathbf{k}} &\approx
	\left\langle \hat{a}^\dag_{\mathbf{k}+\mathbf{q}} \hat{a}_{\mathbf{k}} \right\rangle \hat{a}^\dag_{\mathbf{k}'} \hat{a}_{\mathbf{k}'+\mathbf{q}} +
	\left\langle \hat{a}^\dag_{\mathbf{k}'} \hat{a}_{\mathbf{k}'+\mathbf{q}} \right\rangle \hat{a}^\dag_{\mathbf{k}+\mathbf{q}} \hat{a}_{\mathbf{k}} \\
	&\phantom{{}\approx} - \left\langle \hat{a}^\dag_{\mathbf{k}+\mathbf{q}} \hat{a}_{\mathbf{k}'+\mathbf{q}} \right\rangle \hat{a}^\dag_{\mathbf{k}'} \hat{a}_{\mathbf{k}} 
	- \left\langle \hat{a}^\dag_{\mathbf{k}'} \hat{a}_{\mathbf{k}} \right\rangle \hat{a}^\dag_{\mathbf{k}+\mathbf{q}} \hat{a}_{\mathbf{k}'+\mathbf{q}} \\
	&\phantom{{}\approx} + \left\langle \hat{a}^\dag_{\mathbf{k}+\mathbf{q}} \hat{a}^\dag_{\mathbf{k}'} \right\rangle \hat{a}_{\mathbf{k}'+\mathbf{q}} \hat{a}_{\mathbf{k}}
	+ \left\langle \hat{a}_{\mathbf{k}'+\mathbf{q}} \hat{a}_{\mathbf{k}} \right\rangle \hat{a}^\dag_{\mathbf{k}+\mathbf{q}} \hat{a}^\dag_{\mathbf{k}'}.
\end{align*}
The first line of the right hand side gives the Hartree direct energy with $\mathbf{q}=0$. The second line accounts for the Fock exchange energy, with $\mathbf{k}=\mathbf{k'}$. The third line is the Bogoliubov term, with $\mathbf{k}+\mathbf{q}=-\mathbf{k}'$. With the above approximation, we obtain the Hartree-Fock-Bogoliubov (HFB) Hamiltonian,
\begin{equation}
	H = \int \frac{{\rm d}^3\mathbf{k}}{(2\pi)^3} \left\{ \left[ \frac{|\mathbf{k}|^2}{2m} + U_d + U_e(\mathbf{k}) \right] \left( \hat{a}^\dag_{\mathbf{k}}\hat{a}_{\mathbf{k}} + \hat{a}^\dag_{-\mathbf{k}}\hat{a}_{-\mathbf{k}} \right) + \Delta(\mathbf{k}) \hat{a}^\dag_{\mathbf{k}}\hat{a}^\dag_{-\mathbf{k}} +  \Delta^*(\mathbf{k}) \hat{a}_{-\mathbf{k}}\hat{a}_{\mathbf{k}} \right\},
\end{equation}
where we have defined,
\begin{align}
	U_d &= N \tilde{V}(0), \nonumber \\
	U_e(\mathbf{k}) &= - \int \frac{{\rm d}^3\mathbf{k}'}{(2\pi)^3} \tilde{V}(\mathbf{k}-\mathbf{k}') \langle \hat{a}^\dag_{\mathbf{k}'}\hat{a}_{\mathbf{k}'} \rangle, \nonumber \\
	\Delta(\mathbf{k}) &= \int \frac{{\rm d}^3\mathbf{k}'}{(2\pi)^3} \tilde{V}(\mathbf{k}-\mathbf{k}') \langle \hat{a}_{-\mathbf{k}}\hat{a}_{\mathbf{k}} \rangle, \nonumber
\end{align}
and $N=\int {\rm d}^3\mathbf{k}/(2\pi)^3 \langle \hat{a}^\dag_{\mathbf{k}}\hat{a}_{\mathbf{k}} \rangle $ is the number of particles. The Hartree term, $U_d$, is a constant, and can be absorbed into the chemical potential $\mu$ in a grand canonical ensemble. The Fock term, $U_e(\mathbf{k})$, can modify the kinetic energy. When $U_e(\mathbf{k})$ is anisotropic, it can modify symmetry of the FS. The  gap function $\Delta(\mathbf{k})$ gives the BCS order parameter. It also reveals the Fermi surface instability due to the formation of Cooper pairs. It is antisymmetric with respect to momentum $\mathbf{k}$, i.e. $\Delta(\mathbf{k})=-\Delta(-\mathbf{k})$. By letting $\Delta_\mathbf{k}=0$, one returns to the Hartree-Fock (HF) approximation, where the Fermi surface is well defined by solving $\mu=|\mathbf{k}|^2/(2m) + U_e(\mathbf{k})$. 

With the BCS wave function $
|G\rangle_{\rm BCS} = \prod_{\mathbf{k}} \left( u_\mathbf{k} + v_\mathbf{k} \hat{a}^\dag_{\mathbf{k}} \hat{a}^\dag_{-\mathbf{k}} \right) |0\rangle$, 
it can be found that $\langle \hat{a}^\dag_{\mathbf{k}}\hat{a}_{\mathbf{k}} \rangle =\tilde{\rho}(\mathbf{k})=|v_{\mathbf{k}}|^2$, and $\langle \hat{a}_{-\mathbf{k}}\hat{a}_{\mathbf{k}} \rangle = u^*_{\mathbf{k}}v_{\mathbf{k}}$. The solution to $(u_\mathbf{k}, v_{\mathbf{k}})$ and the self-consistent gap equation reads
\begin{gather} \label{eq:self-consistent1}
	u_\mathbf{k}^2 = \frac12 \left( 1 + \frac{\xi_\mathbf{k}}{\sqrt{\xi_\mathbf{k}^2 + |\Delta_\mathbf{k}|^2}} \right), \quad
	v_\mathbf{k}^2 = \frac12 \left( 1 - \frac{\xi_\mathbf{k}}{\sqrt{\xi_\mathbf{k}^2 + |\Delta_\mathbf{k}|^2}} \right) ,\\ \label{eq:self-consistent2}
	\Delta_\mathbf{k} = \int \frac{{\rm d}^3\mathbf{k}'}{(2\pi)^3} \tilde{V}(\mathbf{k}-\mathbf{k}') \frac{\Delta_{\mathbf{k}'}}{\sqrt{\xi_{\mathbf{k}'}^2 + |\Delta_{\mathbf{k}'}|^2}},
\end{gather}  
where $\xi_{\mathbf{k}}=|\mathbf{k}|^2/(2m) + U_e(\mathbf{k}) - \mu$. The deformation of the FS is associated with the formation of Cooper pairs, which can be quantified by the pair density $\rho_p = \sum_\mathbf{k} u_{\mathbf{k}}^2 v_{\mathbf{k}}^2$. 

The numerical solutions to the momentum density $\tilde{\rho}(\mathbf{k})$ and the order parameter $\Delta(\mathbf{k})$ are obtained by the self-consistent iteration with the Krasnosel'sk\u{\i}-Mann algorithm~\cite{s_Woods2019}. For each iteration step, $\tilde{\rho}(\mathbf{k})$ and $\Delta(\mathbf{k})$ are updated by Eq.~\eqref{eq:self-consistent1}-\eqref{eq:self-consistent2}, and then the new solution is damped by the old one, such that $\tilde{\rho}(\mathbf{k})^{\rm new} = \tilde{\rho}(\mathbf{k})^{\rm new} + \eta [\tilde{\rho}(\mathbf{k})^{\rm old}-\tilde{\rho}(\mathbf{k})^{\rm new}]$. In practice we have used $\eta=0.5$ in the iteration, which provides an efficient way to carry out the numerical simulation. The same iteration procedure is applied to the gap function $\Delta(\mathbf{k})$.

\begin{figure}
	\centering
	\includegraphics[width=0.5\linewidth]{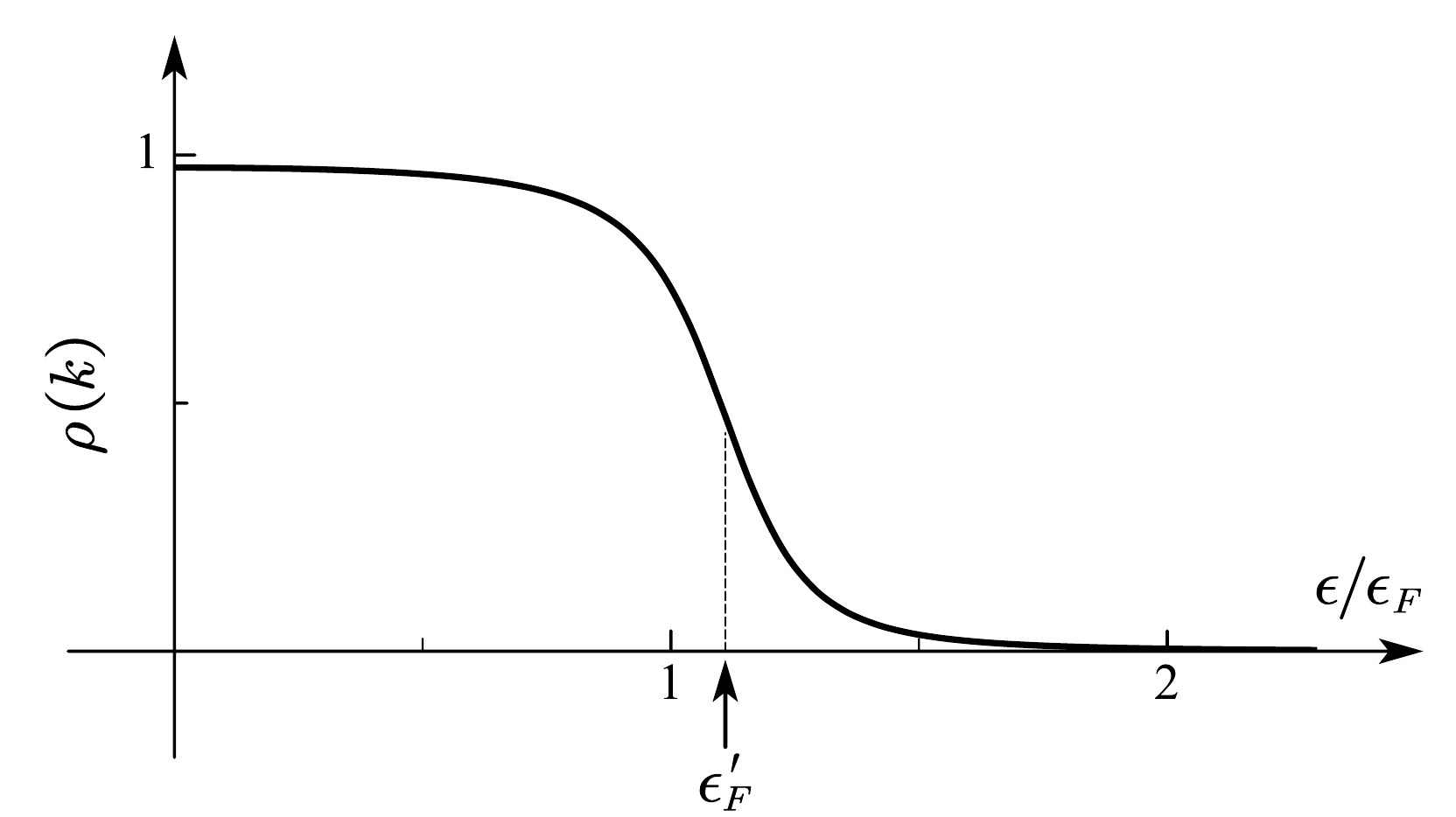}
	\caption{Demonstration of the fitting of the FS.  The fitting parameter $\epsilon_F'$ indicates the energy where the slope of $\rho(k)$ is largest. In this demonstration,  the parameters are  $\epsilon'_F=1.1\epsilon_F$ and $\Delta'_F=0.2\epsilon_F$. }
	\label{fig:Fit_FS}
\end{figure}

\section{Fitting the Fermi surface}\label{appendix:fitting}
In the HFB approximation, the density distribution is depleted, see Fig.~(2) in the main text. Here we employ a fitting procedure to approximately identify the FS~\cite{s_Alexandrov2003}. With the numerical solution of the momentum distribution, we fit the distribution with the following function~\cite{s_Alexandrov2003}
\begin{equation} \label{eq:Fit_func}
	\rho(k,\theta_k) = \frac{1}{2} \left( 1 - \frac{ \epsilon_k - \epsilon'_F(\theta_k) }{ \sqrt{ [\epsilon_k - \epsilon'_F(\theta_k)]^2 + \Delta'(\theta_k)^2 } } \right),
\end{equation}
where $\epsilon'_F(\theta_k)$ and $\Delta'(\theta_k)$ are angular dependent fitting parameters. The parameter $\epsilon'_F(\theta_k) = [k'_F(\theta_k)]^2/2m$ is used to approximately identify the FS, even though the FS is not well-defined. In Fig.~\ref{fig:Fit_FS}, we demonstrate the location of the approximate FS. It shows that parameter $\epsilon'_F(\theta_k)$ gives roughly the largest slope of the density distribution. 

\section{The projection  coefficient}\label{appendix:calculations}
\begin{figure}
	\centering
	\includegraphics{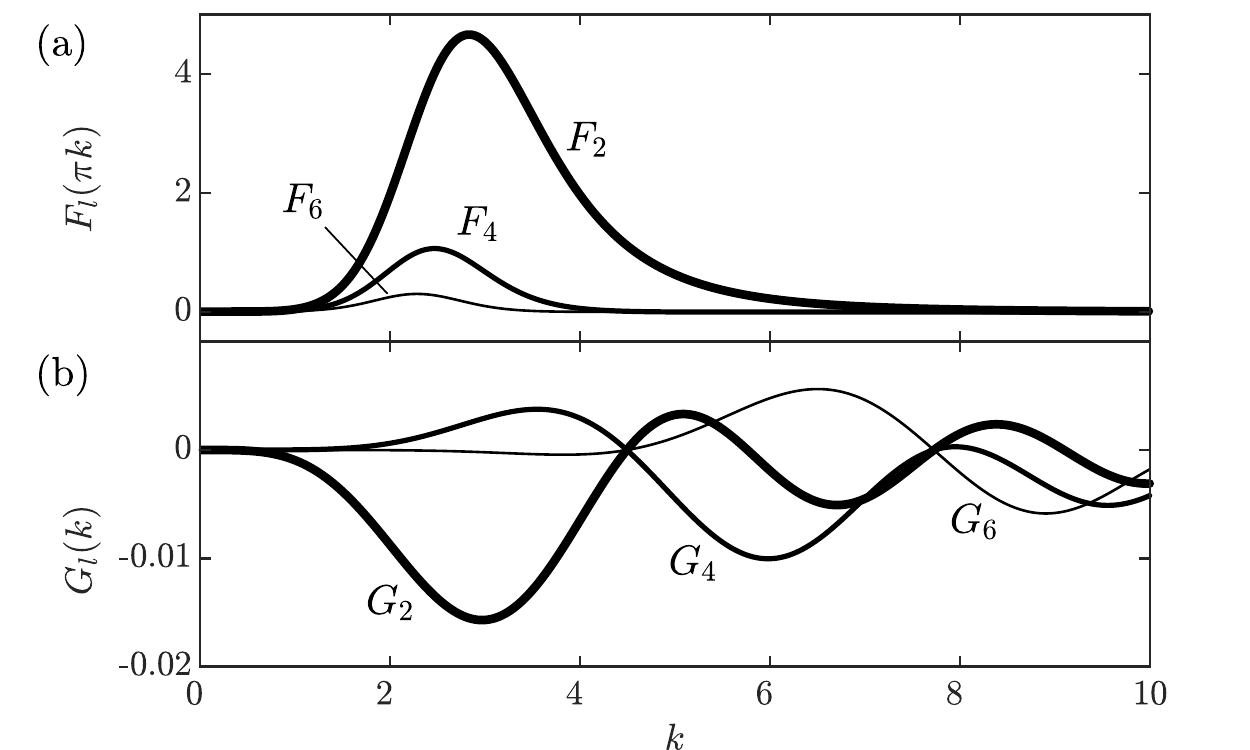}
	\caption{(a) Function $F_l(k)$ has a single peak. For larger $l$, the location of the peak moves towards to lower $k$. (b) Function $G_l(k)$ oscillates with $k$. Increasing $k$, the amplitude of function $G_l(k)$ decreases slowly. }
	\label{fig:perturbationappendix}
\end{figure}
\subsection{Density distribution and the projection coefficient} \label{appendix:perturbation}
In the main text, we showed that the geometry of the deformed FS can be characterized by the projection coefficient $\lambda_l$. Here we illustrate how one can obtain the projection coefficient from the density distribution $\rho(\mathbf{k})$ within the Hartree-Fock approximation. Neglecting the depletion, the density distribution is described approximately  $\tilde{\rho}'(\mathbf{k})=\Theta(k_F+\delta{k}_F(\theta_{k})-|\mathbf{k}|)$. The anisotropy of the approximate FS is characterized by $\delta{k}_F(\theta_k)$. We then project the density distribution $\tilde{\rho}'(\mathbf{k})$ to the spherical harmonics,
\begin{align*}
	\lambda^{\rm HF}_l &= \frac{1}{\rho} \int \frac{{\rm d}^3\mathbf{k}}{(2\pi)^3} \tilde{\rho}'(\mathbf{k}) Y_{l0}(\theta_k,0),\,\, \text{for}\,\, l>0 \\
	&= \frac{1}{k_F^3/(6\pi^2)} \iint \frac{k^2{\rm d}k {\rm d}\Omega_k}{(2\pi)^3}  \Theta(k_F+\delta{k}_F - k(\theta_k)) Y_{l0}(\theta_k,0) \\
	&= \frac{3}{4 \pi k_F^3} \int {\rm d}\Omega_k \int_0^{k_F+\delta{k}_F(\theta_k)} k^2 {\rm d}k Y_{l0}(\theta_k,0) \\
	&= \frac{1}{4 \pi k_F^3} \int {\rm d}\Omega_k \left[ k_F+\delta{k}_F(\theta_k) \right]^3 Y_{l0}(\theta_k,0) \\
	&\approx \frac{1}{4 \pi} \int {\rm d}\Omega_k \left( 1 + 3 \frac{\delta{k}_F(\theta_k)}{k_F} \right) Y_{l0}(\theta_k,0) \\
	&\approx \frac{3}{4\pi} \int {\rm d}\Omega_k \sum_{l'} \frac{1}{2E_F} \lambda'_{l'} Y_{l'0}(\theta_k,0) Y_{l0}(\theta_k,0) \\
	&\approx \frac{3}{8\pi} \frac{\lambda'_l}{E_F}.
\end{align*}
Here $\lambda'_l$ is introduced as we directly expand $\delta{k}_F$,
\begin{equation}
	\frac{\delta{k}_F(\theta_k)}{k_F} = \frac{1}{2E_F} \sum_{l} \lambda'_l Y_{l0}(\theta_k,0).
\end{equation}
The above calculation establishes the relation between $\lambda_l^{\rm HF}$ with $\lambda'_l$. Note that the zeroth term $\lambda_{0}=\sqrt{\frac{1}{4\pi}}$ characterizes the spherical FS, and hence makes no contribution to the FS deformation.

\begin{figure}[b]
	\centering
	\includegraphics{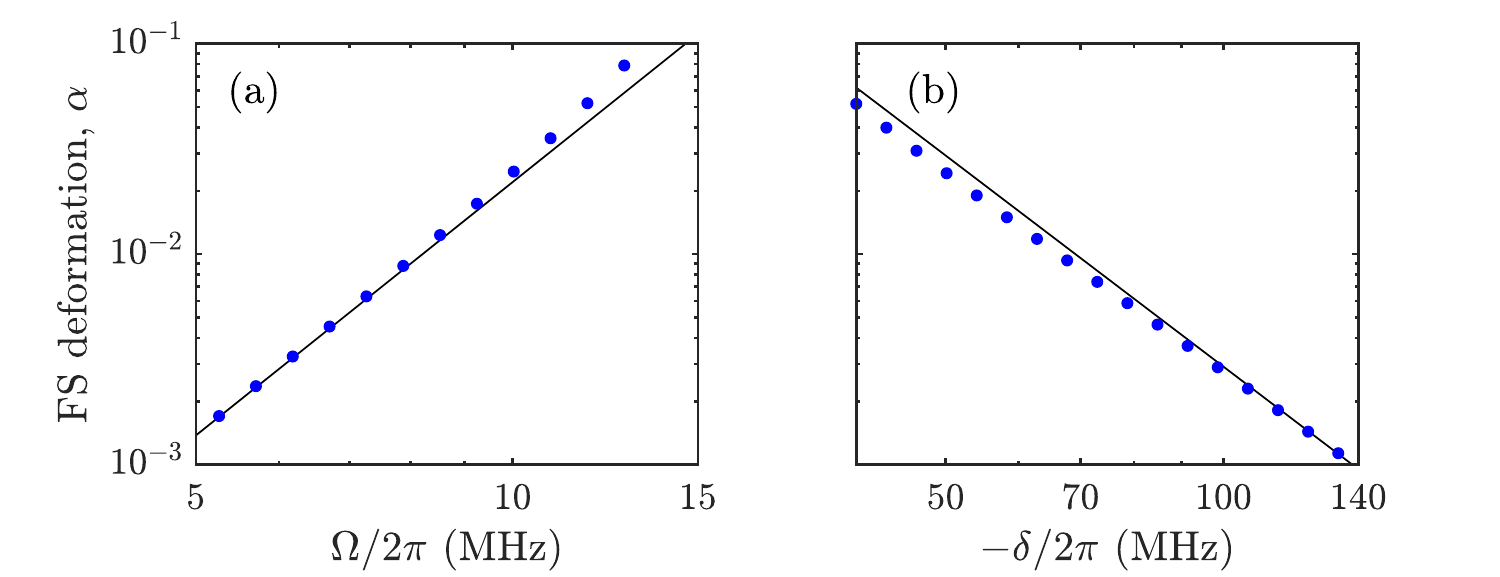}
	\caption{ Scaling relations of the Fermi surface deformation $\alpha$ with respect to (a) $\Omega$ and (b) $\delta$. The quantum number $n=32$ and the initial Fermi vector $k_F=2~{\rm \mu m}^{-1}$, which corresponds to $\bar{R}=1.77~{\rm \mu m}$ and $\rho=1.35\times10^{11}~{\rm cm}^{-3}$. }
	\label{fig:scaling}
\end{figure}
Same derivation can be carried to calculate the projection coefficient with the density distribution $\tilde{\rho}(\mathbf{k})$ obtained from the HBF calculation as follows,
\begin{equation} \label{eq:Lambda_proj}
	\lambda_l = \frac{1}{\rho} \int \frac{{\rm d}^3\mathbf{k}}{(2\pi)^3} \tilde{\rho}(\mathbf{k}) Y_{l0}(\theta_k,0) = \frac{2\pi}{\rho} \int_0^\infty {\rm d} k_z \int_0^\infty {\rm d}k_x \ k_x \tilde{\rho}(k_x,k_z) Y_{l0}\left(\arctan\frac{k_x}{k_z},0\right),
\end{equation}
where $\int {\rm d} k_z$ and $\int {\rm d} k_x$ can be evaluated using the trapezoidal integration method. Similar to the perturbation calculation shown above, one finds the relation, $\lambda_l= \frac{3}{8\pi} \frac{\lambda'_l}{E_F}$ ($l>0$), where $\lambda_l$ is shown in Fig.~3 in the main text.

\subsection{Analysis of the projection coefficient in the perturbative calculation} \label{appendix:scaling}
We now turn to the discussion on the projection coefficient $\lambda_l$. As shown in the main text, the perturbed Fermi surface is given by the convolution of the interaction $\tilde{V}(\mathbf{k})$ and the initial density distribution of the non-interacting Fermi gas, $\tilde{\rho}_0(\mathbf{k}) = \Theta(k_F-|\mathbf{k})$,
\begin{align} \label{eq:Perturbation_Supp}
	\frac{\delta \mathbf{k}_F}{k_{F}} &= \frac{1}{2E_{F}} \int \frac{{\rm d}^3\mathbf{q}}{(2\pi)^3} \tilde{V}(\mathbf{k}_{F}-\mathbf{q}) \tilde{\rho}_{0}(\mathbf{q}) 
	= \frac{1}{2E_{F}} \int {\rm d}^3 \mathbf{r} e^{{\rm i}\mathbf{k}\cdot\mathbf{r}} V(\mathbf{r}) \rho_0^{\rm inv}(\mathbf{r}) \nonumber \\
	&= \frac{1}{2E_{F}} \int r^2\sin\theta {\rm d}r {\rm d}\theta {\rm d}\phi \ V(r,\theta) \rho_0^{\rm inv}(r)  \sum_{l=0,1,2,\dots} 4\pi {\rm i}^l j_l(|\mathbf{k}|r) \sum_{m=-l}^l Y_l^m(\theta,\phi) Y_l^{m*} (\theta_k,\phi_\mathbf{k}) \nonumber \\
	&= \frac{1}{2E_{F}} \sum_{l=0,2,\dots} \int r^2 {\rm d}r \left( 8\pi^2 \int \sin\theta {\rm d}\theta V(r,\theta) Y_{l0}(\theta) \right) \rho_0^{\rm inv}(|\mathbf{k}|r) {\rm i}^l j_l(|\mathbf{k}|r) Y_{l0}(\theta_k) \nonumber \\
	&= \frac{1}{2E_{F}} \sum_{l=0,2,\dots} Y_{l0}(\theta_k) \int {\rm d}k F_l[k/(k_F\bar{R})] G_l(k) \nonumber \\
	&= \frac{1}{2E_{F}} \sum_{l=0,2,\dots} \lambda'_l(k_F\bar{R}) Y_{l0}(\theta_k),
\end{align}
where $\rho_0^{\rm inv}(r) = (2\pi^2)^{-1}j_1(k_Fr)$  is the inverse Fourier transform of $\tilde{\rho}_0(k)$.
The dimensionless functions $F_l(k)$ and $G_l(k)$  are defined by
\begin{subequations}\begin{align}
		F_l(k) &= 8\pi^2 \int_0^\infty \sin\theta{\rm d}\theta \frac{C_6(\theta)/(2\delta \bar{R}^6)}{C_6(\theta)/(2\delta \bar{R}^6)+k^6} Y_{l0}(\theta) , \\
		G_l(k) &= {\rm i}^l \rho_0^{\rm inv}(r) j_l(k). 
\end{align}\end{subequations}
As shown in Fig.~\ref{fig:perturbationappendix}, the function $F_l(k)$ has a single maximum as a function of $k$. $G_l(k)$ oscillates between positive and negative values when increasing $k$, while the amplitude decreases. When calculating the projection coefficient, it turns out that the maximal value of $\lambda_l$ is achieved when the peak regions of $F_l(k)$ and $G_l(k)$ overlap.

In the limit $k_F\bar{R}\ll 1$ ($k_F\ll \bar{k}$), one can make a Taylor's expansion of $\lambda_l$ in terms of $k_F\bar{R}$. The leading order of the expansion gives $\lambda_l \sim V_0/E_F  (k_F \bar{R})^{\beta_l}$ with $\beta_l=l+3$.  Recalling that $V_0= \hbar\Omega^4/8\delta^3$, and $\bar{R}\propto n^{11/6}|\delta|^{-1/6}$, the dependence of the projection on other parameters can be obtained,  $\lambda_l \propto [n^{11\beta_l/6},\Omega^4,|\delta|^{-3-\beta_l/6}, k_F ^{\beta_l-2}]$. In the main text, the dependence on Rydberg state ($n$ and $\bar{R}$) and Fermi momentum (atomic density) is shown in Fig.~3(b)-(c). In Fig.~\ref{fig:scaling}, the dependence of $\lambda_2$ on laser parameter $\Omega$ and $\delta$ is given, which follows the scaling well in the perturbative regime.

\end{document}